\documentclass[submission,publicdomain,hidelinks]{eptcs}
\providecommand{\event}{REFINE 2018} 

\usepackage{latexsym}
\newcommand{\mybox}{\Box}

\usepackage[fleqn]{amsmath}
\newcommand{\mymath}[1]{$ \ #1 \ $}
\newcommand{\mydef}{\textnormal{\Large \ \ \ = \ \ }}
\newcommand{\mywedge}{\ \wedge \ }
\newcommand{\myvee}{\ \vee \ }
\usepackage{bm}
\newcommand{\myrightarrow}{\Large\textbf{$\Rightarrow$}}
\newcommand{\myleftarrow}{\Large\textbf{$\Leftarrow$}}
\newcommand{\myproofdef}{\textnormal{\Large =} \hspace{1em} }
\newcommand{\myproofnodef}{\textnormal{\Large \hphantom{=}} \hspace{1em} }

\title{A Theory of Lazy Imperative Timing}
\author{\href{http://www.cs.utoronto.ca/~hehner}{Eric C.R. Hehner}
\institute{Department of Computer Science\\
University of Toronto, Canada}
\email{\href{mailto:hehner@cs.utoronto.ca}{hehner@cs.utoronto.ca}}
}
\def\titlerunning{A Theory of Lazy Imperative Timing}
\def\authorrunning{Eric Hehner}
\begin{document}
\maketitle

\begin{abstract}
We present a theory of lazy imperative timing.
\end{abstract}

\section{Introduction}

Lazy evaluation was introduced as a programming language execution strategy in
1976 by Peter Henderson and Jim Morris \cite{henderson1976lazy}, and by David
Turner \cite{turner1979new}, and is now part of several programming languages,
including Gofer, Miranda, and Haskell.  It was introduced into the setting of
functional programming, and has mainly stayed there, although it is just as
applicable to imperative programs \cite{guttmann2008lazy}.  The name “lazy
evaluation” is appropriate in the functional setting, but in the imperative
setting it is more appropriately called “lazy execution”.

The usual, familiar execution of programs is called ``eager execution''.  For
example,

\mymath{x:= 2;} \mymath{y:= 3;} \mymath{print \ y}

\noindent
is executed by first executing the assignment \mymath{x:= 2}, and then the
assignment \mymath{y:= 3}, and then the \mymath{print} statement.  If this is
the entire program, a lazy execution executes only the assignment \mymath{y:=
3}, and then the \mymath{print} statement, because the assignment \mymath{x:= 2}
is unnecessary.

Here is a more interesting example.  Let \mymath{i} be an integer variable, and
let \mymath{fac} be an infinite array of integers.

\mymath{i:= 0; \ fac(0):= 1; \ \textbf{while} \ true \ \textbf{do}
\ i:= i + 1; \ fac(i):= fac(i - 1) \times i \ \textbf{od}; \ print fac(3)}

\noindent
After initializing  \mymath{i}  to  \mymath{0}  and  \mymath{fac(0)}  to
\mymath{1}, there is an infinite loop that assigns  \mymath{i!}  (\mymath{i}
factorial) to each array element  \mymath{fac(i)}.  Then, after the infinite
loop, the value of  \mymath{fac(3)}  is printed.  An eager execution will
execute the loop forever, and the final printing will never be done.  A lazy
execution executes only the first three iterations of the loop, and then prints
the desired result.  Of course it is easy to modify the program so that the loop
is iterated only 3 times in an eager execution:  just replace  \mymath{true}  by
\mymath{i < 3}.  But \cite{hughes1989functional} gives a reason for writing it
as above:  to separate the producer (initialization and loop) from the consumer
(printing).  Many programs include a producer and a consumer, and each may be
complicated, and it is useful to be able to write them separately.  When written
as above, we can change the consumer, for example to  \mymath{print fac(4)},
without changing the producer.  It is not the purpose of this paper to argue the
relative merits of eager and lazy execution, nor to advocate any particular way
of programming.  The example is intended to show only that lazy execution can
reduce execution time, and in the extreme case, it can be reduced from infinite
time to finite time.

The analysis of eager execution time is well known;  for example, see
\cite{hehner2012practical}.  Some analysis of lazy execution time has also been
done in the functional setting \cite{sands1990complexity}.  The purpose of this
paper is to present a theory for the analysis of lazy execution time in the
imperative setting.  This paper is based on part of the PhD thesis of Albert Lai
\cite{lai2013eager}, but simplifications have been made to shorten the
explanations, and a different measure of time is being used.

\section{A Practical Theory of Programming}

In \href{http://www.cs.utoronto.ca/~hehner/aPToP}{a Practical Theory of
Programming} \cite{hehner2012practical}, we do not specify programs;  we specify
computation, or computer behavior.  The free variables of the specification
represent whatever we wish to observe about a computation, such as the initial
values of variables, their final values, their intermediate values, interactions
during a computation, the time taken by the computation, the space occupied by
the computation.  Observing a computation provides values for those variables. 
When you put the observed values into the specification, there are two possible
outcomes:  either the computation satisfies the specification, or it doesn't. 
So a specification is a binary (boolean) expression.  If you write anything
other than a binary expression as a specification, such as a pair of predicates,
or a predicate transformer, you must say what it means for a computation to
satisfy a specification, and to do that formally you must write a binary
expression anyway.

A program is an implemented specification.  It is a specification of computer
behavior that you can give to a computer to get the specified behavior.  I also
refer to any statement in a program, or any sequence or structure of statements,
as a program.  Since a program is a specification, and a specification is a
binary expression, therefore a program is a binary expression.  For example, if
the program variables are  \mymath{x}  and  \mymath{y}, then the assignment
program  \mymath{x:= y+1}  is the binary expression  \mymath{x' = y+1 \mywedge
	y'= y}  where unprimed variables represent the values of the program variables
before execution of the assignment, and primed variables represent the values of
the program variables after execution of the assignment.

We can connect specifications using any binary operators, even when one or both
of the specifications are programs.  If  \mymath{A}  and  \mymath{B}  are
specifications, then  \mymath{A \Rightarrow B} says that any behavior satisfying
 \mymath{A}  also satisfies  \mymath{B}, where  \mymath{\Rightarrow}  is
implication.  This is exactly the meaning of refinement.  As an example, again
using integer program variables  \mymath{x}  and  \mymath{y},

\mymath{x := y+1 \ \ \ \ \myrightarrow \ \ \ x'>y}

\noindent We can say ``\mymath{x:= y+1}  implies  \mymath{x'>y}'', or
``\mymath{x:= y+1}  refines  \mymath{x'>y}'', or ``\mymath{x:= y+1}  implements 
\mymath{x'>y}''.  When we are programming, we start with a specification that
may not be a program, and refine it until we obtain a program, so we may prefer
to write

\mymath{x'>y \ \ \ \ \myleftarrow \ \ \ x:= y+1}

\noindent
using reverse implication (``is implied by'', ``is refined by'', ``is
implemented by'').

\section{Eager Timing}

If we are interested in execution time, we just add a time variable \mymath{t}.
Then  \mymath{t}  is the start time, and  \mymath{t'}  is the finish time, which
is  \mymath{\infty}  if execution time is infinite.  We could decide to account
for the real time spent executing a program.  Or we could decide to measure time
as a count of various operations.  In \cite{hehner2012practical} and
\cite{lai2013eager}, time is loop iteration count.  In this paper, time is
assignment count;  I make this choice to keep my explanations short, but I could
choose any other measure.

Using the same program variables  \mymath{x}  and  \mymath{y}, and time
variable  \mymath{t}, the empty program  \mymath{ok}  (elsewhere called
\mymath{skip}), whose execution does nothing and takes no time, is defined as

\mymath{ok \mydef x'=x \mywedge y'=y \mywedge t'=t}

\noindent
An example assignment is

\mymath{x:= y+1 \mydef x'=y+1 \mywedge y'=y \mywedge t'=t+1}

\noindent
The conditional specifications are defined as

\mymath{\textbf{if} \ a \ \textbf{then} \ b \ \textbf{else} \ c \ \textbf{fi}
\mydef a \wedge b \myvee \neg a \wedge c \mydef (a \Rightarrow b) \mywedge (\neg
a \Rightarrow c)}

\mymath{\textbf{if} \ a \ \textbf{then} \ b \ \textbf{fi} \mydef \textbf{if} \ a
\ \textbf{then} \ b \ \textbf{else} \ ok \ \textbf{fi}}

\noindent
A conditional specification is a conditional program if its parts are programs.
The sequential composition  \mymath{A;B}  of specifications  \mymath{A}  and
\mymath{B}  is defined as

\mymath{A;B \mydef \exists x'', y'', t'' \cdot} \hspace{1em} \ (for \ 
\mymath{x', y', t'} \ substitute \ \mymath{x'', y'', t''} \ in \ \mymath{A})

\hspace{9em} \mymath{\mywedge} (for \ \mymath{x, y, t} \ substitute \
\mymath{x'', y'', t''} \ in \ \mymath{B})

\noindent
Sequential composition of  \mymath{A}  and  \mymath{B}  is mainly the
conjunction of  \mymath{A}  and  \mymath{B}, but the final state and time of
\mymath{A}  are identified with the initial state and time of  \mymath{B}.  A
sequential composition is a program if its parts are programs.

In our example program

\mymath{i:= 0; \ fac(0):= 1; \ \textbf{while} \ true \ \textbf{do}
\ i:= i + 1; \ fac(i):= fac(i - 1) \times i \ \textbf{od}; \ print fac(3)}

\noindent
to prove that the execution time is infinite, there are two parts to the proof. 
The first is to write and prove a specification for the loop.  Calling the
specification  \mymath{loop}, we must prove

\mymath{loop \ \ \ \ \myleftarrow \ \ \ \ i:= i+1; \ fac(i):= fac(i - 1) \times
	i; \ loop}

\noindent
The specification we are interested in is

\mymath{loop \mydef t'=t+ \infty}

\noindent
The proof uses\footnote{The arithmetic used here is defined in complete detail
in \cite[p.233-234]{hehner2012practical}.} \mymath{\infty +1 = \infty}, and is
trivial, so we omit it.  If we were to try the specification

\mymath{loop \mydef t'=t+n}

\noindent
for any finite number expression \mymath{n}, the proof would fail because
\mymath{n+1 \neq n}.  A stronger specification that succeeds is

\mymath{loop \mydef (\forall j \le i \cdot  fac'(i)=fac(j)) \mywedge (\forall
j>i\cdot  fac'(j) = fac(i) \times j! \ / \ i!) \mywedge t'=t+ \infty}

\noindent
but the final values of variables after an infinite computation are normally not
of interest (or perhaps not meaningful).  The proof uses  \mymath{(i+1)! = i!
\times (i+1)} and is otherwise easy, so we omit it.

The other part of the eager timing proof is to prove

\mymath{t'=t+ \infty \ \ \ \ \myleftarrow \ \ \ \ i:= 0; \ fac(0):= 1; \ loop; \
	print fac(3)}

\noindent
This proof is again trivial, and omitted.  Eager execution is presented in great
detail in \cite{hehner2012practical}, and is not the point of this paper.

\section{Need Variables}

To calculate lazy execution time, we introduce a time variable and need
variables.  For each variable of a basic (unstructured) type, we introduce a
binary (boolean) need variable.  If  \mymath{x}  is an integer variable, then
introduce binary need variable  \mymath{\mybox x}  (pronounced ``need
\mymath{x}'').  The value of  \mymath{x}  may be  \mymath{7}, or any other
integer;  the value of  \mymath{\mybox x}  may be  \mymath{true}  or
\mymath{false}.  As always, we use  \mymath{x}  and  \mymath{x'}  for the value
of this integer variable at the start and end of a program (which could be a
simple assignment, or any composition of programs).  Likewise we use
\mymath{\mybox x}  and  \mymath{\mybox x'}  for the value of its need variable
at the start and end of a program.  At the start,  \mymath{\mybox x}  means that
the initial value of variable  \mymath{x}  is needed, either in the computation
or following the computation, and  \mymath{\neg \mybox x}  means that the
initial value of variable  \mymath{x}  is not needed for the computation nor
following the computation.  At the end,  \mymath{\mybox x'}  means that the
final value of variable  \mymath{x}  is needed for something following the
computation, and  \mymath{\neg \mybox x'}  means that the final value of
variable  \mymath{x}  is not needed.

With program variables  \mymath{x}  and  \mymath{y}  and time variable 
\mymath{t}, we earlier defined

\mymath{ok \mydef x'=x \mywedge y'=y \mywedge t'=t}

\noindent
We now augment this definition with need variables.  From  \mymath{x'=x}  we see
that the initial value of  \mymath{x}  is needed if and only if the final value
is needed.  Likewise for  \mymath{y}.  So

\mymath{ok \mydef x'=x \mywedge y'=y \mywedge t'=t \mywedge \mybox x=\mybox x'
	\mywedge \mybox y=\mybox y'}

\noindent
Although  \mymath{=}  is a symmetric operator, making  \mymath{x'=x}  and 
\mymath{x=x'}  equivalent, as a matter of style we write \mymath{x'=}(some
expression in unprimed variables) \ because the final value of a program
variable is determined by the initial values of the program variables.  But we
write \mymath{\mybox x=}(some expression of primed need variables) \ because the
need for an initial value is determined by the need for final values.

We now augment the assignment

\mymath{x:= 3 \mydef x'=3 \mywedge y'=y \mywedge t'=t+1}

\noindent
with need variables.  We have a choice.  Perhaps the most reasonable option is

\mymath{x:= 3 \mydef \textbf{if} \ \mybox x' \ \textbf{then} \ x'=3 \mywedge
t'=t+1 \ \textbf{else} \ t'=t \ \textbf{fi} \mywedge y'=y \mywedge \neg \mybox x
\mywedge \mybox y=\mybox y'}

\noindent
This says that if the value of  \mymath{x}  is needed after this assignment,
then that value is  \mymath{3}  and the assignment takes time  \mymath{1}, but
if the value of  \mymath{x}  is not needed afterward, then no final value of 
\mymath{x}  is stated and the assignment takes time  \mymath{0}  because it is
not executed.  In either case, the value of  \mymath{y}  is unchanged.  The
initial value  \mymath{x}  does not appear, so it is not needed, hence
\mymath{\neg \mybox x}.  The last conjunct says that the initial value of 
\mymath{y}  is needed if and only if the final value of  \mymath{y}  is needed,
because  \mymath{y}  appears in the right side of  \mymath{y'=y}.

The other option is

\mymath{x:= 3 \mydef x'=3 \mywedge y'=y \mywedge t'=t+ \textbf{if} \ \mybox x' \
	\textbf{then} \ 1 \ \textbf{else} \ 0 \ \textbf{fi} \mywedge \neg \mybox x
	\mywedge \mybox y=\mybox y'}

\noindent
This option seems less reasonable because it says the final value of  \mymath{x}
 is  \mymath{3}  even if that value is not needed and the assignment is not
executed.  But if the final value of  \mymath{x}  is not used, then it doesn't
hurt to say it's  \mymath{3}.  This option has the calculational advantage that
it untangles the results from the timing.  So this is the option we choose. 
Every assignment has this same timing part, but using the need variable for the
variable being assigned.

In the assignment

\mymath{x:= x+y \mydef x'=x+y \mywedge y'=y \mywedge t'=t+ \textbf{if} \ \mybox
	x' \ \textbf{then} \ 1 \ \textbf{else} \ 0 \ \textbf{fi} \mywedge \mybox
	x=\mybox x' \mywedge \mybox y=(\mybox x' \myvee \mybox y')}

\noindent
we see that  \mymath{x}  appears once, to obtain  \mymath{x'}, so \mymath{\mybox
	x=\mybox x'}.  And  \mymath{y}  appears twice, to obtain \mymath{x'}  and 
\mymath{y'}, so  \mymath{\mybox y=(\mybox x' \myvee \mybox y')}. Time and need
variables can be added automatically, but the algorithm to add them is not
presented in this short paper.

For each structured variable in a program, there is a need variable structured
exactly the same way.  For example, if  \mymath{x: [int, int]}  is a pair of
integers, then  \mymath{\mybox x: [bin, bin]}  is a pair of binaries (booleans);
 the value of  \mymath{\mybox x}  is  \mymath{[true, true]}  or  \mymath{[true,
false]}  or  \mymath{[false, true]}  or  \mymath{[false, false]}.  And if 
\mymath{y}  is an integer variable, then

\mymath{x(0):= 2 \mydef} \hspace{0em} \mymath{x'(0)=2 \mywedge x'(1)=x(1) \mywedge
y'=y \mywedge t'=t+ \textbf{if} \ \mybox x'(0) \ \textbf{then} \ 1 \ \textbf{else}
\ 0 \ \textbf{fi}}

\hspace{7em} \mymath{\mywedge \neg \mybox x(0) \mywedge \mybox x(1)=\mybox x'(1)
\mywedge \mybox y=\mybox y'}

\mymath{x(0):= x(1) \mydef} \hspace{0em} \mymath{x'(0)=x(1) \mywedge x'(1)=x(1)
\mywedge y'=y \mywedge t'=t+ \textbf{if} \ \mybox x'(0) \ \textbf{then} \ 1 \
\textbf{else} \ 0 \ \textbf{fi}}

\hspace{8em} \mymath{\mywedge \neg \mybox x(0) \mywedge \mybox x(1)=(\mybox x'(0)
\myvee \mybox x'(1)) \mywedge \mybox y=\mybox y'}

\mymath{x(0):= y \mydef} \hspace{0em} \mymath{x'(0)=y \mywedge x'(1)=x(1) \mywedge
y'=y \mywedge t'=t+ \textbf{if} \ \mybox x'(0) \ \textbf{then} \ 1 \ \textbf{else}
\ 0 \ \textbf{fi}}

\hspace{7em} \mymath{\mywedge \neg \mybox x(0) \mywedge \mybox x(1)=\mybox x'(1)
\mywedge \mybox y=(\mybox x'(0) \myvee \mybox y')}

If we define datatype  \mymath{tree}  recursively as

\mymath{tree \mydef [ \ ] \ | \ [tree, int, tree]}

\noindent
then a tree is either the empty list, or it is a list of three components, the
first component being the left subtree, the middle component being the root
value, and the last component being the right subtree.  This requires us to
define datatype

\mymath{\mybox tree \mydef bin \ | \ [\mybox tree, bin, \mybox tree]}

\noindent
for need variables.  If we have variable  \mymath{x}  of type  \mymath{tree},
we also have need variable  \mymath{\mybox x}  of type  \mymath{\mybox tree}. 
If  \mymath{x= [ \ ]}, then  \mymath{\mybox x}  is either  \mymath{true}  or 
\mymath{false}.  If  \mymath{x=[[ \ ], 3, [[ \ ], 5, [ \ ]]]}, then 
\mymath{\mybox x}  may be  \mymath{[true, true,[true, true, true]]}  or 31 other
values.

Returning to integer variables  \mymath{x}  and  \mymath{y}, here is an example
conditional program.

\noindent
\hspace{3em} \mymath{\textbf{if} \ x=0 \ \textbf{then} \ y:= 0 \ \textbf{else} \
x:= 0 \ \textbf{fi}}

\noindent
\mymath{\mydef} \hspace{0em} \ \mymath{x'=\textbf{if} \ x=0 \ \textbf{then} \ x \
\textbf{else} \ 0 \ \textbf{fi} \mywedge y'=\textbf{if} \ x=0 \ \textbf{then} \ 0
\ \textbf{else} \ y \ \textbf{fi}}

\noindent
\hspace{3em} \mymath{\mywedge t'=t+\textbf{if} \ x=0 \mywedge \mybox y' \
\textbf{then} \ 1 \ \textbf{else} \ \textbf{if} \ x \neq 0 \mywedge \mybox x' \
\textbf{then} \ 1 \ \textbf{else} \ 0 \ \textbf{fi} \ \textbf{fi}}

\noindent
\hspace{3em} \mymath{\mywedge \mybox x=(\mybox x' \myvee \mybox y') \mywedge \mybox
y=\mybox y'}

\noindent
We see that  \mymath{x}  occurs in the right sides of both  \mymath{x'}  and
\mymath{y'}, so  \mymath{\mybox x=(\mybox x' \myvee \mybox y')}.  We see that
\mymath{y}  occurs in the right side of only  \mymath{y'}, so  \mymath{\mybox
y=\mybox y'}.  We have added the need variables in accordance with the rules,
as we would expect a compiler to do.  But we can do better by using some
algebra.  Notice that  \mymath{\textbf{if} \ x=0 \ \textbf{then} \ x \
\textbf{else} \ 0 \ \textbf{fi} = 0}, so the results part can be stated
equivalently as

\mymath{x'=0 \mywedge y'=\textbf{if} \ x=0 \ \textbf{then} \ 0 \ \textbf{else} \
	y \ \textbf{fi}}

\noindent
which results in the need part

\mymath{\mybox x=\mybox y' \mywedge \mybox y=\mybox y'}

\noindent
We find that  \mymath{\mybox x}  has been strengthened, making lazier execution
possible.  But a compiler would not be expected to make this improvement.

Sequential composition remains the same with need variables added.

\mymath{A;B \mydef \exists x'', y'', t'', \mybox x'', \mybox y'' \cdot}
\hspace{0em} (for \mymath{x', y', t', \mybox x', \mybox y'} substitute
\mymath{x'', y'', t'', \mybox x'', \mybox y''} in \mymath{A})

\hspace{12em} \mymath{\mywedge}(for \mymath{x, y, t, \mybox x, \mybox y}
substitute \mymath{x'', y'', t'', \mybox x'', \mybox y''} in \mymath{B})

At the end of an entire program, we put  \mymath{stop}, defined as

\mymath{stop \mydef x'=x \mywedge y'=y \mywedge t'=t \mywedge \neg \mybox x
	\mywedge \neg \mybox y}

\noindent
Like  \mymath{ok}, its execution does nothing and takes no time.  Since this is
the end of the whole program,  \mymath{\neg \mybox x \mywedge \neg \mybox y}  says
there is no further need for the values of any variables.

\section{Example}

We now have all the theory we need.  Let us apply it to our example program

\mymath{i:= 0; \ fac(0):= 1; \ \textbf{while} \ true \ \textbf{do} \ i:= i + 1;
	\ fac(i):= fac(i - 1) \times i \ \textbf{od}; \ print fac(3); \ stop}

\noindent
To begin, we need a specification for the loop, which we call  \mymath{loop}. 
With a number on each line for reference,

\begin{displaymath}
\begin{array}{llr}

loop \mydef & \ \ \ (\forall j \le i \cdot  fac'(j)=fac(j)) & 0 \\

& \mywedge (\forall j > i \cdot  fac'(j)=fac(i) \times j! / i!) & 1 \\

& \mywedge t'=t +
   \begin{array}[t]{@{}l@{}}
     \textbf{if} \ \mybox i' \ \textbf{then} \ \infty \\ \textbf{else}
     \ \textbf{if} \ \exists j \ge i \cdot \mybox fac'(j) \ \textbf{then} \ 2 \times
     ((max \ j \ge i \cdot  \mybox fac'(j) \cdot  j) - i) \ \textbf{else} \ 0 \
     \textbf{fi} \\ \textbf{fi}
   \end{array}
& 2 \\

& \mywedge \mybox i=(\exists j > i \cdot  \mybox fac'(j)) & 3 \\

& \mywedge (\forall j < i \cdot  \mybox fac(j)=\mybox fac'(j)) & 4 \\

& \mywedge \mybox fac(i)=(\exists j \ge i \cdot  \mybox fac'(j)) & 5 \\

& \mywedge (\forall j > i \cdot  \neg \mybox fac(j)) & 6 \\

\end{array}
\end{displaymath}

Lines 0 and 1 are the same as in the stronger version of the eager specification
presented earlier.  For eager execution lines 0 and 1 are not necessary because
the loop execution is nonterminating, but for lazy execution they are necessary.
 Line 2 is the timing.  It says that if the final value of  \mymath{i}  is
needed, then the loop takes forever; otherwise, if the final value of 
\mymath{fac(j)}  is needed for any  \mymath{j \ge i}, then the loop time is
twice the difference between the largest such \mymath{j}  and  \mymath{i},
because there are two assignments in each iteration;  otherwise the loop takes 
\mymath{0}  time because it will not be executed\footnote{I confess that I did
not get the lazy  \mymath{loop} specification right the first time I wrote it; 
my error was in the time line 2.  The  \mymath{loop} specification is used in
two proofs (below), and any error prevents one of the proofs from succeeding. 
That is how an error is discovered.  Fortunately, the way the proof fails gives
guidance on how to correct the specification.}.  Line 3 says that the loop needs
an initial value for \mymath{i}  if and only if a final value of 
\mymath{fac(j)} is needed for any \mymath{j>i}.  Line 4 says that for 
\mymath{j<i}, \mymath{fac(j)} must have an initial value if and only if its
final value is needed.  Line 5 says that \mymath{fac(i)}  needs an initial value
if and only if the final value of \mymath{fac(j)}  is needed for any  \mymath{j
\ge i}.  And line 6 says that for  \mymath{j>i}, the initial value of
\mymath{fac(j)} is not needed. In this paragraph, the words ``initial'' and
``final'' are used to mean relative to the entire loop execution:  initially
before the first iteration (if there is one), and finally after the last
iteration (if there is one).

There can be more than one specification that's correct in the sense that it
makes the proofs succeed.  For example, if the type of variable  \mymath{i}
allows it, we could add the line  \mymath{i'= \infty}, and then line 3 would be 
\mymath{\mybox i=(\mybox i' \myvee \exists j>i \cdot  \mybox fac'(j))}, but since
line 2 says that if we need  \mymath{i'}  then execution time is infinite, these
additions really don't matter.

The first proof is the loop refinement.  We must prove

\mymath{loop \ \ \ \ \myleftarrow \ \ \ \ i:= i+1; \ fac(i):= fac(i - 1) \times
	i; \ loop}

\noindent
For the proof, we first replace each of the sequentially composed programs with
their binary equivalent, including time and need variables.  Then we use the
sequential composition rule, and use one-point laws to eliminate the
quantifiers.  And we make any simplifications we can along the way.  The proof
is in the Appendix.

Then to prove that the overall execution time is  \mymath{9}, we must prove

\mymath{t'=t+9 \ \ \ \ \myleftarrow \ \ \ \ i:= 0; \ fac(0):= 1; \ loop; \ print
	fac(3); \ stop}

\noindent
For  \mymath{print fac(3)}, we suppose it is like an assignment 
\mymath{print:= fac(3)}, except that  \mymath{print}  is not a program
variable.  This proof is also in the Appendix.

\section{Execution versus Proof}

In a lazy execution, the value of a variable may not be evaluated at various
times during execution.  Nonetheless, the value that would be evaluated if the
execution were eager can still be used in the proof of lazy execution time.  For
example, in the loop specification line 3, we see the conjunct

\mymath{\mybox i=(\exists j>i \cdot \mybox fac'(j))}

\noindent
If there is no \mymath{j>i}  for which  \mymath{fac(j)}  is needed after the
loop, then the value of  \mymath{i}  is not needed before the loop.  The value
of  \mymath{i} is used in the proof to say whether the value of  \mymath{i}  is
needed in execution.

If we change the print statement to  \mymath{print fac(0)}, then the loop is
not executed at all.  The initialization  \mymath{fac(0)}  is still required,
but  \mymath{i:= 0}  is not.  The theory tells us that the execution time is 
\mymath{2}.  The theory still requires that the assignment  \mymath{i:= 0} 
produces  \mymath{i'=0}, but the execution does not.

\section{Conclusion}

We have presented a theory of lazy imperative timing.  The examples presented
are small enough so we know what the right answers are without using the theory;
that enables us to see whether the theory is working.  But the theory is not
limited to small, easy examples.

Time and need variables are added according to a syntactic formula, and that can
be automated.  But in some cases, that formula does not achieve maximum
laziness.  To achieve maximum laziness may require some further algebra.  The
proofs can also be automated, but the prover needs to be given domain knowledge.

\nocite{*}
\bibliographystyle{eptcs}
\bibliography{references}

\appendix
\section{Appendix}

Proof of the loop refinement

\mymath{loop \ \ \ \ \ \myleftarrow \ \ \ \ i:= i+1; \ fac(i):= fac(i-1) \times
	i; \ loop}

\noindent
starting with the right side:

\mymath{i:= i+1; \ fac(i):= fac(i-1) \times i; \ loop}

\begin{flushright}
Replace each statement by its definition.
\end{flushright}
\begin{displaymath}
\begin{array}{rl}

\myproofdef & i'=i+1 \\

\mywedge & (\forall j \cdot  fac'(j)=fac(j)) \\

\mywedge & t'=t+1 \\

\mywedge & \mybox i=\mybox i' \\

\mywedge & (\forall j \cdot  \mybox fac(j)=\mybox fac'(j)); \\

\end{array}
\end{displaymath}
\begin{displaymath}
\begin{array}{rl}

\myproofnodef & i'=i \\

\mywedge &  fac'(i)=fac(i-1) \times i \\

\mywedge & (\forall j \neq i \cdot  fac'(j)=fac(j))  \\

\mywedge & t'=t+1 \\

\mywedge & \mybox i=(\mybox i' \myvee \mybox fac'(i)) \\

\mywedge & (\forall j < i - 1 \cdot \mybox fac(j)=\mybox fac'(j)) \mywedge \mybox
fac(i-1)=(\mybox fac'(i) \myvee \mybox fac'(i-1)) \\

\mywedge & \neg \mybox fac(i) \\

\mywedge & (\forall j > i \cdot \mybox fac(j) = \mybox fac'(j)); \\

\end{array}
\end{displaymath}
\begin{displaymath}
\begin{array}{rl}

\myproofnodef & (\forall j \le i \cdot  fac'(j)=fac(j)) \\

\mywedge & (\forall j>i \cdot  fac'(j)=fac(i) \times j! / i!) \\

\mywedge & t'=t+\textbf{if} \ \mybox i' \ \textbf{then} \ \infty \ \textbf{else} \
\textbf{if} \ \exists j \ge i \cdot  \mybox fac'(j) \ \textbf{then} \ 2 \times
((max j \ge i \cdot  \mybox fac'(j) \cdot  j)-i) \ \textbf{else} \ 0 \ \textbf{fi}
\ \textbf{fi} \\

\mywedge & \mybox i=(\exists j>i \cdot  \mybox fac'(j)) \\

\mywedge & (\forall j<i \cdot  \mybox fac(j)=\mybox fac'(j)) \\

\mywedge & \mybox fac(i)=(\exists j \ge i \cdot  \mybox fac'(j)) \\

\mywedge & (\forall j>i \cdot  \neg \mybox fac(j)) \\

\end{array}
\end{displaymath}
\begin{flushright}
Eliminate the first semi-colon.
\end{flushright}
\begin{displaymath}
\begin{array}{rl}

\myproofdef & i'=i+1 \\

\mywedge &  fac'(i+1)=fac(i) \times (i+1) \\

\mywedge & (\forall j \neq i+1 \cdot  fac'(j)=fac(j))  \\

\mywedge & t'=t+2 \\

\mywedge & \mybox i=(\mybox i' \myvee \mybox fac'(i+1)) \\

\mywedge & (\forall j < i \cdot \mybox fac(j)=\mybox fac'(j)) \mywedge \mybox
fac(i)=(\mybox fac'(i+1) \myvee \mybox fac'(i)) \\

\mywedge & \neg \mybox fac(i+1) \\

\mywedge & (\forall j>i+1 \cdot \mybox fac(j)=\mybox fac'(j)); \\

\end{array}
\end{displaymath}
\begin{displaymath}
\begin{array}{rl}

\myproofnodef & (\forall j \le i \cdot  fac'(j)=fac(j)) \\

\mywedge & (\forall j>i \cdot  fac'(j)=fac(i) \times j!/i!) \\

\mywedge & t'=t+\textbf{if} \ \mybox i' \ \textbf{then} \ \infty \ \textbf{else} \
\textbf{if} \ \exists j \ge i \cdot  \mybox fac'(j) \ \textbf{then} \ 2 \times
((max j \ge i \cdot  \mybox fac'(j) \cdot  j)-i) \ \textbf{else} \ 0 \ \textbf{fi}
\ \textbf{fi} \\

\mywedge & ni=(\exists j>i \cdot  \mybox fac'(j)) \\

\mywedge & (\forall j<i \cdot  \mybox fac(j)=\mybox fac'(j)) \\

\mywedge & \mybox fac(i)=(\exists j \ge i \cdot  \mybox fac'(j)) \\

\mywedge & (\forall j>i \cdot  \neg \mybox fac(j)) \\

\end{array}
\end{displaymath}
\begin{flushright}
Eliminate the last semi-colon. This step uses  \mymath{(i+1)! = i! \times
(i+1)}.
\end{flushright}
\begin{displaymath}
\begin{array}{rl}

\myproofdef & (\forall j \le i \cdot  fac'(j)=fac(j)) \\

\mywedge & (\forall j>i \cdot  fac'(j)=fac(i) \times j!/i!) \\

\mywedge & t'=t+\textbf{if} \ \mybox i' \ \textbf{then} \ \infty \ \textbf{else} \
\textbf{if} \ \exists j \ge i \cdot  \mybox fac'(j) \ \textbf{then} 2 \times
((max j \ge i \cdot  \mybox fac'(j) \cdot  j)-i) \ \textbf{else} \ 0 \ \textbf{fi}
\ \textbf{fi} \\

\mywedge & \mybox i=(\exists j>i \cdot  \mybox fac'(j)) \\

\mywedge & (\forall j<i \cdot  \mybox fac(j)=\mybox fac'(j)) \\

\mywedge & \mybox fac(i)=(\exists j \ge i \cdot  \mybox fac'(j)) \\

\mywedge & (\forall j>i \cdot  \neg \mybox fac(j)) \\

\end{array}
\end{displaymath}
\begin{displaymath}
\begin{array}{rl}
\myproofdef & loop \\
\end{array}
\end{displaymath}

\noindent
\newline
Proof of

\mymath{t'=t+9 \ \ \ \ \ \myleftarrow \ \ \ \ i:= 0; \ fac(0):= 1; \ loop; \
	print fac(3); \ stop}

\noindent starting with the right side:

\mymath{i:= 0; \ fac(0):= 1; \ loop; \ print fac(3); \ stop}

\begin{flushright}
Replace each statement by its definition.
\end{flushright}
\begin{displaymath}
\begin{array}{rl}

\myproofdef & i'=0 \\

\mywedge & (\forall j \cdot  fac'(j)=fac(j)) \\

\mywedge & t'=t+\textbf{if} \ \mybox i' \ \textbf{then} \ 1 \ \textbf{else} \ 0
\ \textbf{fi} \\

\mywedge & \neg \mybox i \\

\mywedge & (\forall j \cdot  \mybox fac(j)=\mybox fac'(j)); \\

\end{array}
\end{displaymath}
\begin{displaymath}
\begin{array}{rl}

\myproofnodef & i'=i \\

\mywedge & fac'(0)=1 \\

\mywedge & (\forall j>0 \cdot  fac'(j)=fac(j)) \\

\mywedge & t'=t+\textbf{if} \ \mybox fac'(0) \ \textbf{then} \ 1 \ \textbf{else}
\ 0 \ \textbf{fi} \\

\mywedge & \mybox i=\mybox i' \\

\mywedge & \neg \mybox fac(0) \\

\mywedge & (\forall j>0 \cdot  \mybox fac(j)=\mybox fac'(j)); \\

\end{array}
\end{displaymath}
\begin{displaymath}
\begin{array}{rl}

\myproofnodef & (\forall j \le i \cdot  fac'(j)=fac(j)) \\

\mywedge & (\forall j>i \cdot  fac'(j)=fac(i) \times j!/i!) \\

\mywedge & t'=t+\textbf{if} \ \mybox i' \ \textbf{then} \ \infty \ \textbf{else}
\ \textbf{if} \ \exists j \ge i \cdot  \mybox fac'(j) \ \textbf{then} \ 2 \times
((max j \ge i \cdot  \mybox fac'(j) \cdot  j)-i) \ \textbf{else} \ 0 \
\textbf{fi} \ \textbf{fi} \\

\mywedge & \mybox i=(\exists j>i \cdot  \mybox fac'(j)) \\

\mywedge & (\forall j<i \cdot  \mybox fac(j)=\mybox fac'(j)) \\

\mywedge & \mybox fac(i)=(\exists j \ge i \cdot  \mybox fac'(j)) \\

\mywedge & (\forall j>i \cdot  \neg \mybox fac(j)); \\

\end{array}
\end{displaymath}
\begin{displaymath}
\begin{array}{rl}

\myproofnodef & print=fac(3) \\

\mywedge & i'=i \\

\mywedge & (\forall j \cdot  fac'(j)=fac(j)) \\

\mywedge & t'=t+1 \\

\mywedge & \mybox i=\mybox i' \\

\mywedge & \mybox fac(3) \\

\mywedge & (\forall j \neq 3 \cdot  \mybox fac(j)=\mybox fac'(j)); \\

\end{array}
\end{displaymath}
\begin{displaymath}
\begin{array}{rl}

\myproofnodef & i'= i \\

\mywedge & (\forall j \cdot  fac'(j)=fac(j)) \\

\mywedge & t'=t \\

\mywedge & \neg \mybox i \\

\mywedge & (\forall j \cdot  \neg \mybox fac(j)) \\

\end{array}
\end{displaymath}
\begin{flushright}
Eliminate the semi-colons and simplify.
\end{flushright}
\begin{displaymath}
\begin{array}{rl}

\myproofdef &  print=6 \\

\mywedge & (\forall j \cdot  fac'(j)=j!) \\

\mywedge & t'=t+9 \\

\mywedge & \neg \mybox i \\

\mywedge & (\forall j \cdot  \neg \mybox fac(j)) \\

\end{array}
\end{displaymath}
\begin{flushright}
Use specialization.
\end{flushright}
\begin{displaymath}
\begin{array}{rl}
\myrightarrow \hspace{0.5em} & t'=t+9
\end{array}
\end{displaymath}

\end{document}